
\input jnl
\input reforder
\input defs
\input  defs.tex
\input eqnorder
\rightline{NSF-ITP-95-14}

\title Effective Field Theory

\vskip .2in

\author  A. Zee

\affil
Institute for Theoretical Physics
University of California
Santa Barbara, CA 93106-4030 USA

\vskip .3in

\abstract
{I give a brief review of effective field theory, discussing the
contribution of
Feza G\"ursey in particular and focussing on the literature I am most
familiar
with.}

\vskip .5in

I would like to begin by saying a
few  words about Feza G\"ursey (1921-1992),
whom I
regard as one of the last ``gentleman-physicists." Unlike many of the other
speakers, I never had the pleasure of any direct physics interaction with
him.
But I
have met him on a number of occasions and Feza and Suha have always
been
exceptionally nice to me, ever since the beginning of my career. The last
time
I saw them was a few years ago in a small town in eastern Hungary.
Every time Feza saw me, he told me to come to Istanbul. Well, here I  am
finally, but unfortunately in his absence. The community has lost a true
gentleman scholar.

My subject today is the low energy or long distance
effective field theory, a
concept that has pervaded throughout much of
modern physics. In a sense,
all of physics involves the concept of effective
field theory.
Hydrodynamics, for example, studies the behavior of a
collection of
particles on distance scales large compared to the separation
between the
particles. One can even say that all of known physics may be described by
the effective low effective Lagrangian of string theory. In recent years, the
concept of effective field theory has played an increasingly important role
in
condensed matter physics as well as in particle physics.

Perhaps the two most studied effective Lagrangians are the Landau-
Ginzburg
theory of superconductivity and the sigma model of the interaction between
pions and nucleons. \footnote {}{Lecture given at the G\"ursey Memorial
Conference on Strings and Symmetries, June 1994, to be
published by Springer-Verlag.}\vadjust{\eject}
The Landau-Ginzburg theory went on to
great glory as
the prototype of a spontaneously broken gauge theory which underlies
electroweak unification and grand unification. For its part, the non-linear
sigma model has been studied intensively in recent years in connection with
quantum spin systems, both ferromagnetic and anti-ferromagnetic. The
discovery of high
temperature superconductivity have thrust these studies into prominence
since the relevant materials are known to be
anti-ferromagnetic at low doping concentrations. In these applications, it is
the non-linear sigma model, rather than the linear sigma model, that enters.

The ubiquitous non-linear sigma model first appeared in the work of Feza
G\"ursey.\refto{g1, g2, g3, chg} Indeed, even the notation and the
philosophical underpinning in G\"ursey's first paper\refto{g1} were
already
remarkably close to what is used in modern times. Starting with the chiral
transformation of the nucleon field
$$
\psi (x)\rightarrow e^{2i
\gamma_{5} {\vec \tau} . {\vec \theta} }\psi (x)
$$
G\"ursey jumped to the
non-linear transformation
$$
\psi (x) \rightarrow e^{2i \gamma_{5}
{\vec \tau} . {\vec \phi (x)} }\psi
(x)
$$
where $\phi$ denoted the pion
field. Incidentally, G\"ursey cited
Nishijima\refto{nishi} for this crucial
step of replacing the parameter of a
symmetry transformaton, ${\vec
\theta}$, by a field, ${\vec \phi (x)} $. He
then
identified the unitary
matrix field
$
\Phi = e^{2 i f {\vec \tau} . {\vec \phi}}
$, its kinetic
energy term
$tr \partial \Phi \partial \Phi^{\dagger}$
and its interaction
with nucleons. As another
historical note, we mention here that G\"ursey
cited Glauber\refto{glauber} as
having written down, in 1951, a non-linear interaction of pions with
nucleons in order to account for multiple
pion production in nucleon
nucleon
collisions. Of course, back in 1951,
chiral invariance was not yet
appreciated.

Shortly afterwards, indeed,
in the same volume of {\it Il Nuovo
Cimento},
Gell-Mann and
L\'evy\refto{gml} wrote down the linear sigma model. They were able to
make the theory renormalizable, at
the
cost of introducing another field,
the sigma field. The meson described by the sigma field, while its existence
was, and remains controversial, has managed to give its name to this class
of
field theories. To put all
this into perspective, we must remember that
in the early sixties,
renormalizability was considered ``sacred," and non-renormalizable field
theories, such as the non-linear sigma model, were
regarded with distaste.
More on this later.

Indeed, at that time even the
relevance of field theory for strong
interaction
physics was much in
doubt, and the emphasis was decidely on the S-matrix
and the dispersion
theoretic approach. The notion of broken chiral
invariance
was
established only with the successes of the current algebra
approach\refto{addashen} championed by Gell-Mann and others in the
mid
1960's. Using current algebra Adler and Weisberger were able to
calculate
what amounts to the low-energy interaction between pions and
nucleons, an
approach developed further by a number of authors. In
particular,
Weinberg\refto{pionscatter}
showed how to calculate pion pion scattering at low
energies, and in effect
re-discovered the non-linear sigma model. In a
series of influential papers,
Weinberg not only brought respectability back
to the non-linear sigma
model, but to the entire Lagrangian field theory
approach, thus sweeping
away the S-matrix worship of the late fifties and
early sixties and paving
the
way for electroweak unification.

In the
early seventies, these considerations were extended to the interaction
between pions and photons. Indeed, Schwinger's
calculation of the pion decaying into two photons
amounts
to finding the effective Lagrangian at energies low compared to
the
nucleon
mass, and in this sense represents an intellectual descendant
of Euler and
Heisenberg's calculation of the effective four-photon
interaction at energies
low compared to the electron mass. Adler, B. Lee,
Treiman, and I, without
ever mentioning the word Lagrangian or the
word field, used various
consistency requirements to
determine\refto{altz, terentiev, earlier}  the amplitudes for the
processses $\gamma
\rightarrow 3 \pi$ and for $2 \gamma \rightarrow 3
\pi$. At the same
time, and completely independently, Wess and
Zumino\refto{wess} found
the effective Lagrangian describing these
processes. The Lagrangian
appeared quite strange: its action can only be
written as a five dimensional
integral.

Again, to put things into perspective, we should recall that the
emphasis at
that period in the history of particle physics was on
momentum-space
amplitudes, on the differential cross sections\refto{zee}
that actually could
be measured at various newly built electron
accelerators. An effective
Lagrangian was regarded as only a mnemonic
device. Independently of
Wess
and Zumino, but during and shortly after
their work had already appeared,
Aviv and I used Schwinger's proper
time method\refto{propertime} to determine the effective
Lagrangian describing the interaction of an arbitrary number of $SU(3)$
octet mesons and photons.\refto{aviv} Lacking the insight of Wess and
Zumino, we wrote out the Lagangian as an infinite series. These days this
exercise would be referred to as evaluating the fermion determinant.

Since
as an intermediate step we had to work out the quark propagator in the
presence of external meson and electromagnetic fields, we could close the
fermion line to obtain not only the effective Lagrangian, but also other
fermion bilinears as well. In particular, Aviv and I also wrote down the
effective current.\refto{current}  Some years later, Goldstone and
Wilczek\refto{gw} rediscovered that
using
the matrix field $\Phi$ introduced by G\"ursey (with the Pauli matrices
$\vec
\tau$ promoted to the $SU(3)$ Gell-Mann matrices $\vec \la$) the effective
baryon current
could be
written in a compact form
$$
J ^{\mu}= \hbox{constant}\; \epsilon^{\mu\nu\la\sigma} tr
(\Phi^{\dagger}\part_{\nu} \Phi \Phi^{\dagger}\part_{\la} \Phi
\Phi^{\dagger}\part_{\sigma} \Phi)
$$

\refis{current} Aviv and I calculated the electromagnetic current $\bar \psi
Q \gamma_{\mu} \psi$ with $Q$ the charge matrix of the quark fields
$\psi$;
there was no reason to consider the baryon current at that time. In our
final result (equations (4.44) and (4.46) in \Ref {aviv}) the matrix $Q$
can
be simply set to unity for comparison with Goldstone-Wilczek.

Aviv and I used another representation of the non-linear sigma model,
discussed by G\"ursey and by Gell-Mann and L\'evy,  in which
$\sigma$ and $\varphi$ were constrained by $\sigma^2 + \varphi^2 = f^2$,
and
thus failed to see the invariant group structure made apparent in the
Goldstone-Wilczek form. Various strands in the development of effective
Lagrangian and effective currents were all interconnected in an interesting
way.\refto{az}

\refis{az} A. Zee, \pl 135B, 307, 1984.

Years later, Witten\refto{witten} developed the subject further, starting
with
the five-meson scattering amplitude and realizing that this could not be
described by a Lagrangian in four-dimensional spacetime. Nowadays, this
can be seen quite easily by noting that in analogy with the effective  current
written above, the effective Lagrangian would have the form $\cL \sim
\epsilon^{\mu\nu\la\sigma\tau} tr (\Phi^{\dagger}\part_{\mu}
\Phi\Phi^{\dagger}\part_{\nu} \Phi \Phi^{\dagger}\part_{\la} \Phi
\Phi^{\dagger}\part_{\sigma} \Phi \Phi^{\dagger}\part_{\tau} \Phi)$ and
an
antisymmetric symbol with five indices is available only in
five-dimensional
spacetime. The five-pion amplitude is forbidden by G-parity, but in $SU(3)
\times
SU(3)$ five-meson scattering is allowed. Witten brought the theory into the
modern form.

As mentioned earlier, the non-linear sigma model has been extensively
studied in
connection with ferromagnetism and anti-ferromagnetism in the condensed
matter physics literature.
Naively, it seems reasonable enough that in a ferromagnetic or
anti-ferromagnetic system the magnetization (or the staggered
magnetization)
can be represented in the continuum limit by a 3-vector field ${\vec
n}(x, t)$. We are typically not interested in the fluctuation of the
magnitude of ${\vec n}(x, t)$ and thus ${\vec n}(x, t)$ may be taken
to be a vector of unit length. The Lagrangian is naturally taken to be a
non-linear sigma model
$$
\cL = {1\over 2f^2} \part_{\mu} {\vec n} . \part^{\mu} {\vec n}
$$

For a long time, a number
of field theorists puzzled over how quantum spin, even a single quantum
spin with its non-commuting components, could be incorporated into the
path integral formalism. I understand
that Feynman himself was troubled over this point. In hindsight, as if often
the case, the solution as now presented in textbooks\refto{fradkin} seems
straightforward enough, and it appears as if one only needs to have ``one's
head screwed on straight" to be led by the formalism step-by-step towards
the correct answer, namely that the Wess-Zumino-Witten term has to be
included.

In fact, there is another approach which avoids having to write the action
as a higher dimensional integral. To explain this, I must regrettably
mention
that particle physicists were traditionally confused by the difference
between the ferromagnet and the anti-ferromagnet, as indicated by the
discussion in the paragraph preceding the one above. A clue is
provided by the fact, as shown in
elementary solid state physics texts, that the spin wave disperses linearly
(that is, $\omega \propto k$) in an antiferromagnet and quadratically (that
is, $\omega \propto k^2$) in a ferromagnet. Thus, on the face of it, the
ferromagnet cannot be represented by the (relativistic) non-linear sigma
model written above. This suggests that the Lagrangian has to involve only
one derivative in time, rather than two, but there is no such term involving
$\vec n$! The term $\vec n . \part_t \vec n$ is a total derivative.

The solution,\refto{wespin} as it turns out, was to write $\vec n =
z^{\dagger} \vec \sigma z$ with $z$ a spinor such that $z^{\dagger}  z=1$.
The desired  Lagrangian  is then
$$
\cL = -i z^{\dagger} \dot z - V(\vec n)
$$
with  $V(\vec n)=
(\nabla \vec n)^2\; +\;$ possible other terms such as the coupling of $\vec
n$
to an external magnetic field. Indeed, using the identity
$\delta (z^{\dagger} \dot z) \propto \delta \vec n
(\vec n \times \dot {\vec n})
$
we obtain
$$
\vec n \times \dot {\vec n} = {{\delta V}\over {\delta \vec n}}
$$
Taking the scalar product of this equation with $\vec n$ we obtain
$$
\dot {\vec n} = \vec n \times {{\delta V}\over {\delta \vec n}}
$$
the familiar equation of motion for a spin. A straightforward
exercise\refto{wespin}
shows that the dispersion laws around a ferromagnetic and an
anti-ferromagnetic background are indeed different and as stated above.

\refis{fradkin} E. Fradkin, {\it Field Theories of Condensed Matter
Systems}, (Addison-Wesley Publishing Co. 1991).

\refis{wespin} X. G. Wen and A. Zee, \prl 61, 1025, 1988.

My generation of physicists was taught the notion that quantum field
theory,
quantum electrodynamics, quantum chromodynamics, and so forth, were
``fundamental,"  that these theories
hold
at arbitrarily short distance scales. We used renormalization group to study
the ultraviolet flow towards short distances. We would write down a
Lagrangian and use ``renormalizability" and symmetry to limit the number
of
possible terms. A ``more modern" view, which has emerged from
condensed
matter physics, in particular the theory of critical phenomena, and which
may be called ``Wilsonian", acknowledges and emphasizes that the short
distance physics may be extremely complicated or unknown, or perhaps
even ``unknowable," depending on your philosophical persuasion. In
condensed matter physics, the short distance physics may be described by
lattice dynamics. In particle physics, the short distance physics is allegedly
that of a string. Instead of the ultraviolet flow, we should study the
infrared
flow towards long distances and times and hope that most terms become
irrelevant in that limit. We would then arrive at an effective field theory
described by a small number of terms. The better among the more modern
textbooks, such as the one by Polyakov, specifically emphasize this view.
The
relevant equations are essentially the same, but the mindset is different.

We are thus instructed to study the renormalization group flow of various
operators. In many cases, simple dimensional analysis, which may be
regarded as ``zeroth order renormalization group," suffices. In particle
physics, this point of view was developed over a number of years. An early
example is a ``model independent" analysis of proton decay.\refto{sw, wz}

Let us say that we believe only in $SU(3) \times SU(2) \times U(1)$ and
not in
grand unification. We simply say that proton decay is due to some
unknown short distance physics, but whatever this mechanism might be, we
can still write down a long distance effective
Lagrangian to describe proton decay. The most relevant operators are
those with the lowest scaling dimensions. These dimension six operators
have the form
$$
{1\over M_*^2} qqql
$$
where $q$ and $l$ represent quark and lepton fields respectively. By
engineering dimensional analysis, an unknown mass $M_*$ has to be
introduced. Thus, the rate for proton decay is of course undetermined.
However, by imposing $SU(3) \times SU(2) \times U(1)$ we are able to
restrict the
number of possible operators enormously. In this way, we can make
predictions about proton decay completely independent of what the short
distance physics may be!

Incidentally, this may be construed as an argument for colored quarks.
Before the invention of quarks, we were able to write down a dimension
four operator of the form $P{\bar e}\pi$, and we must arbitrarily decree
that the dimensionless coefficient in front of this operator to be
ridiculously small, and since we don't understand why this coefficient is so
small, we dignify this ignorance by elevating it to a principle, known as
baryon conservation. It is color that forces us to go from a dimension four
operator to a dimension six operator. Physics has progressed: a small
dimensionless coefficient has been replaced by the ratio $($proton
mass$)^2/M_*^2$. We can go on and study any exotic process involving
the
known quarks and leptons, by systematically writing down,\refto{weldon}
in
order of increasing scaling dimension, all possible operators allowed by
$SU(3) \times SU(2) \times U(1)$.

Given an effective long distance field theory, we can of course try to
``induce" what the short distance physics might be. In general, of course,
many possible short distance physics may give rise to the same long
distance physics: this remark has been elevated to the ``principle of
universality.Ó Given the long distance physics, we can arrive at the correct
short distance physics only by astutely combining experimental
observations and inspired guesses  guided by general principles and
esthetic
considerations. Such is the  history of physics. Currently, this enterprise is
represented by string theory.

An infinitely more modest example involves a possible Majorana mass for
the neutrino. The relevant long distance effective Lagrangian, as restriced
by $SU(3) \times SU(2) \times U(1)$,  has the form
$$
\cL_{eff} = f \tilde{ \psi_L} C \psi_L \tilde {\phi} \phi
$$
where $\psi_L$ and $\phi_L$ represent the left-handed lepton doublet and
the Higgs doublet respectively. Here {$\tilde \phi$} denotes, as usual,
$\epsilon_{ij} \phi_j$. When the neutral component of $\phi$ acquires a
vacuum expectation
value, the neutrino gets a Majorana mass. Can we ``induce" the short
distance physics responsible? In other words, can we replace effectively the
dimension five
operator above by operators with dimension four or less. One attempt is
represented by the following\refto{neutrino}
$$
\cL = g \tilde {\psi_L} C \psi_L h^+  + M \tilde {\phi} \phi h^-
$$
where $h$ represents a charged $SU(2)$ singlet Higgs. The neutrino
Majorana
mass is now calculable (in the technical sense!) in terms of the (unknown)
$g$ and $M$. Short distance physics of course tells us more (again, the
progress of physics!). In this particular case, we can learn something about
the flavor of the neutrino Majorana mass.\refto{smirnov}

In some sense, physics is possible only if the effect of a particle of mass
$M$ on the low energy or long distance effective theory vanishes as $M$
tends to infinity. Thus, for Schwinger to calculate the $g$ factor of the
electron, he
didn't have to know the mass of the top quark. There is however a
conceptually important exception to this perfectly reasonable expectation.
Suppose that after the field for the massive particle is removed, the
resulting theory becomes non-renormalizable. Then the low energy theory
will remember the massive particle: its absence would be missed even at
arbitrarily low energies.

For example, in the standard model the left handed top quark appears in a
doublet $(t\; b)_L$. Removing the top field would render the theory
non-renormalizable. Now consider\refto{collins} the Feynman diagram in
which a $Z$ boson couples to a top quark loop which connects to an
external up or down quark line via two gluons.  This process introduces an
isoscalar contribution to the neutral current of the up and down quarks. A
simple analysis shows that indeed this radiative correction goes like $\log
m_t$ for large top quark mass $m_t$. Note that whether or not the theory
left behind after the removal of the heavy field is renormalizable or not
depends on the theory being considered. Thus, for Schwinger, the relevant
theory was quantum electrodynamics in which the top quark is represented
as just another charged field. Removing it leaves the theory perfectly
renormalizable.

Another striking example is given by the electric dipole moment of the
neutron\refto{weinberg} and electron.\refto{barr} Consider the diagram
in which a photon couples to a top quark loop which connects to an
external electron line via a photon and a Higgs field. Again, we see that this
contribution to the electric dipole moment of the electron grows like $\log
m_t$. One instructive way of looking at this process is to consider an
effective Lagrangian for the coupling of a Higgs field to two photons,
which in fact consists the dimension five operator $\phi F_{\mu\nu}
F^{\mu\nu}$. When we
insert this effective Lagrangian into a calculation
of the electric dipole moment of the electron, we would obtain a
(logarithmically) divergent answer since this Lagrangian is not
renormalizable. The relevant Feynman integral has to be cut off at the
energy scale beyond
which the low energy effective Lagrangian ceases to be valid, and this scale
is set by precisely the top quark mass.

In condensed matter physics, we are not interested in ultraviolet
divergences since these are always cut off by the lattice. As an example,
consider the chiral spin state\refto{WWZ, fradkin}. Start with a single
particle hopping on a square lattice. We can look at the ``Feynman
freshman physics" book for example and see that the energy of the particle
is related to its momentum by $E=-(cos k_x + cos k_y)$; the spectrum,
typical of band structure theory, reflects the square symmetry of the
lattice. Now suppose every time the electron goes around a square
plaquette
on the lattice it acquires a factor of (-1). Another way of saying this is that
there is a magnetic flux of $\pi$ threading through each plaquette. The
energy spectrum becomes\refto{AM, K, IL} $E = \pm \sqrt {cos^2 k_x +
cos^2
k_y}$,
still a messy looking expression. But if we want to study a half-filled
system, that is, if we fill the band with fermions up to $E=0$ in accordance
with the Pauli exclusion principle, and if we are only interested in physics
at long distance and time, then we expand around a point where $E$
vanishes. Writing $k_x = {\pi \over 2} + q_x$ and $k_y = {\pi \over 2} +
q_y$, we find $E =\pm  \sqrt {q_x^2 + q_y^2}$. This is a most remarkable
result: in a system which is not even rotational invariant we obtain a
Lorentz invariant dispersion!

\refis{AM} I. Affleck and J.B. Marston, \prb 37, 3774, 1988.

\refis{K} G. Kotliar, \prb 37, 3664, 1988.

\refis{IL} L.B. Ioffe and A.I. Larkin, \prb 39, 8988, 1989.

Further, when we allow the particle to hop along the diagonal as well, such
that when the particle hops around a triangle its wave function acqures a
phase of $i$, a gap opens up between the upper and lower band. The
effective low energy theory describing a half-filled system, that is for
energies low compared to the gap, then reads
$$
\cL = {\bar \psi} (i  \slp- m) \psi + .....
$$
We have discovered the Dirac Lagrangian! Finally, introducing the phase
degree of freedom contained in the fermion creation and annihilation
operator on the lattice, we arrive at the gauge theory
$$
\cL = {\bar \psi} (i  \slp + a - m) \psi
$$
For physics below the energy scale $m$, we can integrate out the fermion
field and thus obtain
$$
\cL = {1\over 4\pi} 2 \epsilon^{\mu\nu\la} a_{\mu} \part_{\nu}
a_{\la} + .....,
$$
the famous Chern-Simon term much talked about by physicists and
mathematicians.\refto{chu} From here we can proceed to a discussion of
fractional statistics and semion superconductivity, but that is another
story.\refto{semion} One can also go on to discuss extensions to $3+1$
dimensional spacetime and to non-abelian flux and so on.\refto{nonabel,
tik}

The remarkable emergence of a relativistic Dirac Lagrangian from a lattice
theory without even rotational invariance naturally prompts speculations on
whether the observed quarks and leptons can emerge in this way
also.\refto{begbook} In a
sense, that is what lattice gauge theory\refto{kog} is all about.

We do not want to give the impression that low energy effective theories
are easy to write down. The difficulty is that the relevant degrees of
freedom may be quite different from those present in the short distance
theory. In quantum chromodynamics, for example, we have hadrons in the
low energy theory, not quarks.
In superconductivity, we have the Ginzburg-Landau field or the Cooperon.
In the example just given, the relevant low energy degree of freedom turns
out to be a gauge potential governed by Chern-Simons dynamics.

A striking recent example of this phenomenon is the effective field theory
of quantum Hall fluid, which after all, simply consists of electrons
interacting and moving in a $2+1$ dimensional spacetime with a
perpendicular magnetic field. The microscopic physics is described by a
trial wave function of the $N \sim 10^{23}$ electrons proposed by
Laughlin. Here I cannot possibly give you anything more than just the
flavor of the effective field theory approach.

What I will do here is to argue what the
effective field theory must be from general principles.\refto{FLCH} The
conservation of the electromagnetic current $\part^\mu
J_\mu=0$ in (2+1)-dimensional spacetime tells us that
the current can be
written  as the curl
of a vector potential, then reads
$$J^\mu={1\over 2\pi}\eps^{\munu\la} \part_\nu
a_\la $$ We now note that when we transform  $a_\mu$ by
$a_\mu\to
a_\mu-\part_\mu\La$,  the current is unchanged.
We did not go looking for a gauge potential, the gauge potential came
looking for us! There is no place to hide. The  existence of a
gauge potential follows from completely general
considerations.

Let us now try to write down a low energy effective local
Lagrangian in terms of operators of lowest possible dimensions, noting that
parity and time reversal are broken by the external magnetic field. Since
gauge
invariance
forbids the dimension 2 term $a_\mu a^\mu$ in the
Lagrangian,
the
simplest possible term is in fact the dimension 3 Chern-Simons
term
$\eps^{\munu\la}a_\mu\part_\nu a_\la$. Thus, the Lagrangian
(density) is
simply
$$\cL={k\over 4\pi}\eps^{\munu\la}a_\mu\part_\nu a_\la+\ldots$$
To determine the dimensionless
coefficient  $k$ we
couple the system to an ``external'' or ``additional''
electromagnetic gauge
potential $A_\mu$, thus
now have
$$\cL={k\over 4\pi}
\eps^{\mu\nu\la} a_\mu \part_\nu a_\la + {1\over
2\pi}
\eps^{\munu\la} A_\mu
\part_\nu a_\la={k\over 4\pi}
\eps^{\munu\la} a_\mu \part_\nu a_\la
-{1\over 2\pi} \eps^{\munu\la}
a_\mu\part_\nu A_\la $$
Integrating out  $a$ we obtain $$\cL_{\rm eff}
=-
{1\over 4\pi k} \eps^{\munu\la} A_\mu\part_\nu A_\la.
$$
The electromagnetic current that flows in response to
$A_\mu$ is thus
$$J^\mu\equiv -{\de\cL_{\rm eff}
\over\de A_\mu}
={1\over 2\pi k} \eps^{\munu\la}
\part_\nu A_\la $$
Looking at the time component of this equation, we recognize $1/k$ as the
ratio of the density of electrons to the magnetic field. To study the
elementary excitations in the system, we simply couple the current of these
excitations to the gauge potential $a$. Proceeding in this
way, we can easily recover the classic Laughlin results that the elementary
excitations carry fractional charge and statistics. For details
and references to the original literature, I refer the reader to some
pedagogical lectures I gave last winter.\refto{africa}

\refis{africa}  A. Zee, ``Field theory of quantum Hall fluids'', to appear in
the proceedings of the South African School on Field Theory and
Condensed Matter Systems, Tsitsikamma, South Africa, 1994, to be
published by Springer-Verlag.

Another topic I like to mention is that of tunneling in double-layered
quantum Hall systems. First, in the absence of tunneling between the two
layers, the current associated with each layer is separately conserved $J_I =
{1\over
2\pi} \epsilon \part a_I$,  $I=1,2$. Thus, we can generalize the effective
Lagrangian above to
$$
\cL={1\over 4\pi}( \sum_{IJ} K_{IJ} a \eps \part a +\sum_I 2A\eps \part a
)+
a_1 j_1+a_2 j_2 $$
with a 2-by-2 matrix $K=\pmatrix{l&n\cr n&m\cr}$  We note in
passing that this $K$ matrix is in one-to-one correspondences with a class
of wave functions proposed long ago by Halperin\refto{Hlp} to describe
double-layered Hall systems.

What happens if $K$ has
a zero eigenvalue? Then some linear combination of
the
gauge fields becomes massless, leading to a gapless mode and some
interesting physics.\refto{wzmono1}

 How is tunneling represented in
this picture? When an
electron
tunnels from one layer to another, the currents $J_1^\mu = {1\over
2\pi}\epsilon\part a_1$ and $J_2^\mu= {1\over 2\pi}\epsilon\part a_2$ are
no longer separately conserved. Since electrons are represented by flux
quanta, tunneling from the first layer to the second converts flux of type
$\epsilon\part a_1$ to flux of type  $\epsilon\part a_2$. Thus tunneling
corresponds to a kind of magnetic monopole into which flux of type 1
disappears and out of which flux of type 2 appears. Indeed, more formally,
we have
$$
\Delta(N_1-N_2)=\pm 2 =
\int dt\; d^2x\;\part_\mu
(J_1^\mu -J_2^\mu)={1\over 2\pi}\int dt\; d^2x\;
\part_\mu(\eps^{\mu\nu\la}
\part_\nu a_{-,\la})
$$
Suppose we continue from Minkowskian (2+1) dimensional spacetime to
Enclidean $3$ space. We recognize $\part_\mu(\eps^{\mu\nu\la}
\part_\nu a_{-,\la})$  as
$\vec\nabla\cdot(\vec\nabla\times\vec A)=\vec\nabla\cdot\vec B$
if we identify $a_{-,\lambda}$ as a 3-vector gauge potential $\vec A$ in
Euclidean space and $\vec B$ the corresponding magnetic field.
 This is precisely a Dirac
magnetic monopole, with its flux quantized to be $\pm
4\pi$, just as Dirac said it should be.

Thus, in Euclidean $3$ space we have a plasma of magnetic monopoles and
anti-monopoles. At the location of each monopole and each anti-monopole
there occurs a tunneling event in spacetime. Polyakov\refto{pkv}
showed long ago
that in the presence of a monopole plasma the photon acquires a mass with
an effective  sine-Gordon Lagrangian
$$
\cL_{eff}=g^2(\partial\theta)^2+\zeta\cos\theta
$$

It is immensely pleasing that some of the most profound concepts in
theoretical physics are involved here. The discretness of the electron leads
to Dirac quantization of the magnetic monopoles. The quantization of
monopoles leads to an angular variable as the order parameter.The
appearance of an angular order parameter immediately reminds us of
the Josephson effect in superconductors. Wen and I\refto{wzmono1}
were thus led to predict
that there should be ``Josephson-like" effects in double-layered quantum
Hall systems. We were careful to use the term ``Josephson-like" because the
double-layered Hall system is to be sure not a superconductor, and thus the
usual discussion of the Josephson effect must be taken over here with great
care. A detailed discussion is beyond the scope of these lectures. We refer
to the original work\refto{wzmono1, wzmono2, wzmono3} and to the
recent  literature.\refto{yang,moon,ich}

\refis{ich} I. Ichinose and T. Ohbayashi, U. Tokyo preprint 1994.

An interesting probe is provided by applying
a magnetic field parallel to the plane of the double-layered Hall system.
When an electron tunnels from one layer to another, its wave function
now acquires a phase factor
$$
e^{\pm ie\int dz A_z} \equiv e^{\pm i\xi(x)}
\eqno(tunphase)
$$
(We denote the coordinate perpendicular to the plane by $z$ and the
two-dimensional coordinates in the plane by $x$.) A monopole is now
associated with the
phase factor $e^{+i\xi(x_a)}$, and an anti-monopole with the phase factor
$e^{-i\xi(x_a)}$. It is not difficult to verify that in the effective
Lagrangian the cosine term is now modified to $cos
(\theta + \xi)$. Wen
and I\refto{wzmono2} considered a random magnetic field and showed
that its effect was to reduce the tunneling parameter $\zeta$. Yang et
al\refto{yang} showed that a uniform magnetic field would drive an
interesting commensurate-incommensurate transition.

Again, the non-linear sigma model of G\"ursey appears naturally. Here the
vector field $\vec n$ is an order parameter such that when it is pointing up
it indicates electrons in the upper layer, and when it is pointing down,
electrons in the lower layer. Thus, if we impose a voltage  across the
double layered system we simply add to the Lagrangian
a term $Un_z$ which says that to have an electron in the upper layer
costs more energy than to have it in the lower layer. It is also easy to see
that tunneling can be represented by a field driving $\vec n$ to point in the
$x$-direction (because $S_x = S_+ + S_-$ raises and lowers electrons
between the upper and lower layers.) In this way, we arrive at a more
involved  non-linear sigma model\refto{yang} $$
\cL = -i z^{\dagger} {\dot z} - (\nabla {\vec n})^2 - \beta n_z^2 - \eta
n_x
$$
The sine-Gordon mentioned earlier can be obtained\refto{wzmono3} as an
effective low
energy Lagrangian from this effective Lagrangian by integrating out the
fast mode, namely $n_z$. As is in general the case, there is a hierarchy of
low
energy effective Lagrangians, presumably with string theory at the top of
the hierarchy.

In closing, I would like to mention some recent work\refto{bz1, bz2, bz3}
on random
matrix theory. First, a faster-than-lightning review of this venerable
subject. In the early fifties, Wigner posed the problem of calculating the
distribution of the eigenvalues of an $N$ by $N$ (as $N
\rightarrow \infty$) hermitean matrix  randomly taken from a probability
distribution, for
example,
$$
P(\phi) = {1\over Z} e^{-N tr V(\phi)}
$$
where $V$ is a polynomial of its argument. The operator measuring the
density of eigenvalues is given by $\hat \rho (\mu) = {1\over N} tr
\delta(\mu - \phi)$. The density of eigenvalues $\rho(\mu)=<\hat \rho
(\mu)>$ has been known for some fifteen years,\refto{BIPZ} and as one
might expect, depends on $V$ of course. For $V$ an even polynomial, the
density is non-vanishing between $-a$ and $a$ where the width of the
spectrum $a$ is a complicated functional of $V$.

What  about the density-density correlation function $\rho_c(\mu,
{\nu) =\; <\hat \rho (\mu)\hat \rho (\nu)>\; -\;}\break
{<\hat \rho
(\mu)><\hat \rho (\nu)>}$?
This correlation function was determined recently for any $V$ and was
found to have wild oscillations, as expected since there are $N$ eigenvalues
in the spectrum. It is convenient to think of the eigenvalues as the positions
of a gas of atoms living in a one-dimensional space of width $2a$. The
short distance physics depends on $V$ in detail.

The surprising discovery\refto{bz1, bz3} is that when $\rho_c(\mu,\nu)$
is
smoothed over
these short distance details, it becomes universal when expressed in terms
of the obvious scaling variables $x=\mu/a$, $y=\nu/a$, that is, we found
the
smoothed correlation to be\refto{bz1}
$$
\rho^{{\rm smooth}}_c (\mu,\,\nu)  =
{-1\over2N^2\pi^2 a^2} f(x,y)$$
where the function
$$
f(x,y)={1\over(x-y)^2}
{(1-xy)\over[(1-x^2)(1-y^2)]^{1/2}}.
$$ is universal in the sense that it does not depend on $V$ at all.
We find this result rather remarkable since the density $\rho(\mu)$ does
depend on $V$. Even after the Gaussian law of large numbers has been
proved to us, it
seems perhaps somewhat mysterious that random numbers would ``know"
about the function $e^{-x^2}$. In the same way, it appears mysterious that
somehow random matrices know about the function $f(x,y)$.

Here the long distance effective theory corresponds to hydrodynamics, and
not to a renormalization group flow towards low energy. Indeed, the
universality can be derived using hydrodynamics.\refto{been} Alternate
derivations have also been given.\refto{eynard, forrester}

\refis{been} C. W. J. Beenakker, to appear in {\sl Nucl. Phys. B}.

\refis{eynard} B. Eynard, hep-th/9401165, to appear {\sl Nucl. Phys. B}.

\refis{forrester} P. J. Forrester, to be published.

There is however also an analog\refto{BZJ} of the renormalization group.
In
the
renormalization group  approach, we thin out the number of degrees of
freedom. Here we can take an $N$ by $N$ matrix, integrate over its last
row and column, and obtain an $N-1$ by $N-1$ matrix. In this way, we
can obtain a renormalization group flow to determine the density of
eigenvalues. The calculation\refto{bz2} is particularly
``neat" because none of the usual complications appears.

\vskip .25in
\centerline{\bf ACKNOWLEDGEMENTS}
\vskip .15in

In reading a history of the Turks, I learned that the word  ``Turk'' first
appeared in an ancient Chinese chronicle, which describe the  Turks as
exceptionally hospitable ``milk drinking octonion loving  barbarians.'' I
would
like to thank my Turkish colleagues for their  exceptional  hospitality. This
work is supported in part by the National Science  Foundation under Grant
No.
PHY89-04035.

\references

A note to my colleagues who are not referenced here: I feel that this may
not
be the appropriate place to
give a comprehensive set of references to the recent literature. In my more
extensive review
article\refto{africa} a more complete set of references may be found.

\refis{g1} F. G\"ursey, \journal Nuovo Cimento, 16, 230, 1960.

\refis{g2} F. G\"ursey, \journal Annals of Physics, 12, 91, 1961.

\refis{chg} P. Chang and F. G\"ursey, \pr 164, 1752, 1967.

\refis{gml} M. Gell-Mann and M. L\'evy, \journal Nuovo Cimento, 16, 705,
1960.

\refis{bz1} E. Br\'ezin and A. Zee, \np 402(FS), 613, 1993.

\refis{bz2} E. Br\'ezin and A. Zee, \journal Compt.\ Rend.\
Acad.\
Sci.\/, 317, 735, 1993.

\refis{bz3} E. Br\'ezin  and A. Zee, \pr E49, 2588, 1994.

\refis{addashen} S. Adler and R. Dashen, {\sl Current\ Algebra\/} (W. A.
Benjamin, Inc. New York, 1968).

\refis{altz} S. Adler, B. W. Lee, S. B. Treiman, and A. Zee, \pr D4, 3497,
1971.

\refis{terentiev} Results similar to those of Adler, Lee, Treiman, and Zee
were
given independently by M. V. Terentiev, JETP Letters 14, 140, 1971.

\refis{wess} J. Wess and B. Zumino, \pl 37B, 95, 1971.

\refis{zee} A. Zee, \pr D6, 885, 1972.

\refis{g3} F. G\"ursey, ``Effective Lagrangians in Particle Physics,'' in {\it
Particles, \ Currents, \ and \ Symmetries\/} (Acta Phys. Austriaca, Suppl.
V,
185, 1968).

\refis{glauber} R. J. Glauber, \pr 84, 395, 1951.

\refis{weldon} A. Weldon and A. Zee, \np  173B, 269, 1980.

\refis{BZJ} E. Br\'ezin and J. Zinn-Justin, \pl B288, 54, 1992.




\refis{BIPZ} E. Br\'ezin, C. Itzykson, G. Parisi, and J.B. Zuber,
\journal Comm. Math. Phys., 59, 35, 1978.





\refis{smirnov}  For a review and references to earlier work, see
A. Y. Smirnov and Z. Tao, Trieste preprint 1994.

\refis{collins} J. C. Collins, F. Wilczek, and A. Zee, \pr D18, 242, 1978.

\refis{barr} S. M. Barr and A. Zee, \prl  65, 21, 1990.

\refis{weinberg} S. Weinberg, \prl  63, 2333, 1989.

\refis{nishi} K. Nishijima, \journal Nuovo Cimento, 11, 910, 1959.

\refis{witten} E. Witten, \np B223, 422, 1983.

\refis{aviv} R. Aviv and A. Zee, \pr  D5, 2372, 1972

\refis{gw} J. Goldstone and F. Wilczek, \prl 47, 986, 1981.

\refis{wzmono1} X.G. Wen and A. Zee, \prl 69, 1811, 1992.

\refis{wzmono2} X.G. Wen and A. Zee, \prb 47, 2265, 1993.

\refis{wzmono3} X.G. Wen and A. Zee, ``A Phenomenological Study of
Interlayer Tunnelling in Double-Layered Quantum Hall Systems," MIT-
ITP preprint, 1994.

\refis{FLCH} J. Fr\"ohlich and A. Zee, \np B364, 517, 1991.

\refis{yang} K. Yang, K. Moon, L. Zheng, A. H. MacDaonald, L. Zheng,
D. Yoshioka, and S. C. Zhang, \prl 72, 732, 1994.

\refis{WWZ} X.G. Wen, F. Wilczek, and A. Zee, \pr B39, 11413, 1990.

\refis{chu} {{\it Physics\ and\ Mathematics\ of\ Anyons\/}}, edited by S. S.
Chern \etal, World Scientific Publishing 1990.

\refis{begbook} A. Zee, ``Emergence of spinor from flux and lattice
hopping,"Ó, in {{\it M.\ A.\ B.\ Beg\ Memorial\ Volume\/}}, edited by A.
Ali and P. Hoodbhoy, World Scientific Publishing 1990.

\refis{semion} A. Zee,  ``Semionics'', in {{\it High Temperature
Superconductivity}}, edited by K. Bedell, D. Coffey, D. Pines and J.R.
Schrieffer,
Addison-Wesley 1990.

\refis{kog} J. Kogut and L. Susskind, \pr D11, 395, 1975.

\refis{Hlp} B.I. Halperin, \journal Phys. Helv. Acta., 56, 75, 1983.

\refis{pkv} A. Polyakov, \np B120, 429, 1977.

\refis {moon} K. Moon, H. Mori, K. Yang, S.M. Girvin, A. H.
MacDonald, L. Zheng, D. Yoshioka, and S. C. Zhang, Indiana preprint
1994.

\refis{propertime} J. Schwinger, \pr 82, 664, 1951.

\refis{earlier} The work of Adler, Lee, Treiman, and Zee was inspired by,
but did not agree with, the earlier work of R. Aviv, N. D. Hari Dass, and R. F.
Sawyer, \prl 26, 591, 1971, and of T. Yao, \pl 35B, 225, 1071.

\refis{pionscatter} S. Weinberg, \prl 17, 168, 1966.

\refis{wz} F. Wilczek and A. Zee, \prl 43, 1571, 1979; \pl 88B, 311, 1979.

\refis{sw} S. Weinberg, \prl 43, 1566, 1979.

\refis{nonabel} A. Zee, \journal Int. J. Mod. Physics, B5, 529, 1991, Z.
Kunszt and A. Zee, \pr B44, 6842, 1991.

\refis{tik} A. M. Tikofsky, S. B. Libby, and R. B. Laughlin, \np B413,
579, 1994.

\refis{neutrino} A. Zee, \pl 93B, 389, 1980.

\endreferences

\end


\message
{JNL.TEX version 0.95 as of 5/13/90.  Using AM fonts.}

\catcode`@=11
\expandafter\ifx\csname inp@t\endcsname\relax\let\inp@t=\input
\def\input#1 {\expandafter\ifx\csname #1IsLoaded\endcsname\relax
\inp@t#1%
\expandafter\def\csname #1IsLoaded\endcsname{(#1 was previously loaded)}
\else\message{\csname #1IsLoaded\endcsname}\fi}\fi
\catcode`@=12

\font\twelverm=amr10 scaled 1200    \font\twelvei=ammi10 scaled 1200
\font\twelvesy=amsy10 scaled 1200   \font\twelveex=amex10 scaled 1200
\font\twelvebf=ambx10 scaled 1200   \font\twelvesl=amsl10 scaled 1200
\font\twelvett=amtt10 scaled 1200   \font\twelveit=amti10 scaled 1200
\font\twelvesc=amcsc10 scaled 1200  \font\twelvesf=amssmc10 scaled 1200
\font\twelvemib=ambi10 scaled 1200
\font\tensc=amcsc10                     \font\tensf=amss10
\font\tenmib=ambi10
\font\eightmib=ambi10 scaled 800
\font\sixmib=ambi10 scaled 667
\skewchar\twelvei='177			\skewchar\twelvesy='60
\skewchar\twelvemib='177

\newfam\mibfam

\def\twelvepoint{\normalbaselineskip=12.4pt plus 0.1pt minus 0.1pt
  \abovedisplayskip 12.4pt plus 3pt minus 9pt
  \belowdisplayskip 12.4pt plus 3pt minus 9pt
  \abovedisplayshortskip 0pt plus 3pt
  \belowdisplayshortskip 7.2pt plus 3pt minus 4pt
  \smallskipamount=3.6pt plus1.2pt minus1.2pt
  \medskipamount=7.2pt plus2.4pt minus2.4pt
  \bigskipamount=14.4pt plus4.8pt minus4.8pt
  \def\rm{\fam0\twelverm}          \def\it{\fam\itfam\twelveit}%
  \def\sl{\fam\slfam\twelvesl}     \def\bf{\fam\bffam\twelvebf}%
  \def\mit{\fam 1}                 \def\cal{\fam 2}%
  \def\sc{\twelvesc}		   \def\tt{\twelvett}%
  \def\sf{\twelvesf}               \def\mib{\fam\mibfam\twelvemib}%
  \textfont0=\twelverm   \scriptfont0=\tenrm   \scriptscriptfont0=\sevenrm
  \textfont1=\twelvei    \scriptfont1=\teni    \scriptscriptfont1=\seveni
  \textfont2=\twelvesy   \scriptfont2=\tensy   \scriptscriptfont2=\sevensy
  \textfont3=\twelveex   \scriptfont3=\twelveex\scriptscriptfont3=\twelveex
  \textfont\itfam=\twelveit
  \textfont\slfam=\twelvesl
  \textfont\bffam=\twelvebf \scriptfont\bffam=\tenbf
                            \scriptscriptfont\bffam=\sevenbf
  \textfont\mibfam=\twelvemib \scriptfont\mibfam=\tenmib
                              \scriptscriptfont\mibfam=\eightmib
  \normalbaselines\rm}


\def\tenpoint{\normalbaselineskip=12pt plus 0.1pt minus 0.1pt
  \abovedisplayskip 12pt plus 3pt minus 9pt
  \belowdisplayskip 12pt plus 3pt minus 9pt
  \abovedisplayshortskip 0pt plus 3pt
  \belowdisplayshortskip 7pt plus 3pt minus 4pt
  \smallskipamount=3pt plus1pt minus1pt
  \medskipamount=6pt plus2pt minus2pt
  \bigskipamount=12pt plus4pt minus4pt
  \def\rm{\fam0\tenrm}\def\it{\fam\itfam\tenit}%
  \def\sl{\fam\slfam\tensl}\def\bf{\fam\bffam\tenbf}%
  \def\mit{\fam 1}\def\cal{\fam 2}%
  \def\sc{\tensc}\def\tt{\tentt}%
  \def\sf{\tensf}\def\mib{\fam\mibfam\tenmib}%
  \textfont0=\tenrm   \scriptfont0=\sevenrm   \scriptscriptfont0=\fiverm
  \textfont1=\teni    \scriptfont1=\seveni    \scriptscriptfont1=\fivei
  \textfont2=\tensy   \scriptfont2=\sevensy   \scriptscriptfont2=\fivesy
  \textfont3=\tenex   \scriptfont3=\tenex     \scriptscriptfont3=\tenex
  \textfont\itfam=\tenit
  \textfont\slfam=\tensl
  \textfont\bffam=\tenbf \scriptfont\bffam=\sevenbf
                         \scriptscriptfont\bffam=\fivebf
  \textfont\mibfam=\tenmib \scriptfont\mibfam=\eightmib
                           \scriptscriptfont\mibfam=\sixmib
  \normalbaselines\rm}

\mathchardef\alpha="710B
\mathchardef\beta="710C
\mathchardef\gamma="710D
\mathchardef\delta="710E
\mathchardef\epsilon="710F
\mathchardef\zeta="7110
\mathchardef\eta="7111
\mathchardef\theta="7112
\mathchardef\iota="7113
\mathchardef\kappa="7114
\mathchardef\lambda="7115
\mathchardef\mu="7116
\mathchardef\nu="7117
\mathchardef\xi="7118
\mathchardef\pi="7119
\mathchardef\rho="711A
\mathchardef\sigma="711B
\mathchardef\tau="711C
\mathchardef\phi="711E
\mathchardef\chi="711F
\mathchardef\psi="7120
\mathchardef\omega="7121
\mathchardef\varepsilon="7122
\mathchardef\vartheta="7123
\mathchardef\varpi="7124
\mathchardef\varrho="7125
\mathchardef\varsigma="7126
\mathchardef\varphi="7127


\def\beginlinemode{\endmode
  \begingroup\parskip=0pt \obeylines\def\\{\par}\def\endmode{\par\endgroup}}
\def\beginparmode{\endmode
  \begingroup \def\endmode{\par\endgroup}}
\let\endmode=\par
{\obeylines\gdef\
{}}
\def\singlespace{\baselineskip=\normalbaselineskip}

\def\oneandahalfspace{\baselineskip=\normalbaselineskip
  \multiply\baselineskip by 3 \divide\baselineskip by 2}
\def\doublespace{\baselineskip=\normalbaselineskip \multiply\baselineskip by 2}

\newcount\firstpageno
\firstpageno=2
\footline={\ifnum\pageno<\firstpageno{\hfil}\else{\hfil\twelverm\folio\hfil}\fi}
\def\toppageno{\global\footline={\hfil}\global\headline
  ={\ifnum\pageno<\firstpageno{\hfil}\else{\hfil\twelverm\folio\hfil}\fi}}
\let\rawfootnote=\footnote		
\def\footnote#1#2{{\rm\singlespace\parindent=0pt\parskip=0pt
  \rawfootnote{#1}{#2\hfill\vrule height 0pt depth 6pt width 0pt}}}
\def\raggedcenter{\leftskip=4em plus 12em \rightskip=\leftskip
  \parindent=0pt \parfillskip=0pt \spaceskip=.3333em \xspaceskip=.5em
  \pretolerance=9999 \tolerance=9999
  \hyphenpenalty=9999 \exhyphenpenalty=9999 }
\def\dateline{\rightline{\ifcase\month\or
  January\or February\or March\or April\or May\or June\or
  July\or August\or September\or October\or November\or December\fi
  \space\number\year}}
\def\received{\vskip 3pt plus 0.2fill
 \centerline{\sl (Received\space\ifcase\month\or
  January\or February\or March\or April\or May\or June\or
  July\or August\or September\or October\or November\or December\fi
  \qquad, \number\year)}}


\hsize=6.5in
\hoffset=0pt
\vsize=8.9in
\voffset=0pt
\parskip=\medskipamount
\def\\{\cr}
\twelvepoint		
\doublespace		
\overfullrule=0pt	




\def\title			
  {\null\vskip 3pt plus 0.2fill
   \beginlinemode \doublespace \raggedcenter \bf}

\def\author			
  {\vskip 3pt plus 0.2fill \beginlinemode
   \singlespace \raggedcenter\sc}

\def\affil			
  {\vskip 3pt plus 0.1fill \beginlinemode
   \oneandahalfspace \raggedcenter \sl}

\def\abstract			
  {\vskip 3pt plus 0.3fill \beginparmode
   \oneandahalfspace ABSTRACT: }

\def\endtitlepage		
  {\endpage			
   \body}

\def\body			
  {\beginparmode}		

\def\head#1{			
  \goodbreak\vskip 0.5truein	
  {\immediate\write16{#1}
   \raggedcenter \uppercase{#1}\par}
   \nobreak\vskip 0.25truein\nobreak}

\def\itemitemitem{\par\indent\indent \hangindent3\parindent \textindent}
\def\itemitemitemitem{\par\indent\indent\indent \hangindent4\parindent
\textindent}
\def\beginitems{\par\medskip\bgroup
  \def\i##1 {\par\noindent\llap{##1\enspace}\ignorespaces}%
  \def\ii##1 {\item{##1}}%
  \def\iii##1 {\itemitem{##1}}%
  \def\iiii##1 {\itemitemitem{##1}}%
  \def\iiiii##1 {\itemitemitemitem{##1}}
  \leftskip=36pt\parskip=0pt}\def\enditems{\par\egroup}

\def\makefigure#1{\parindent=36pt\item{}Figure #1}

\def\figure#1 (#2) #3\par{\goodbreak\midinsert
\vskip#2
\bgroup\makefigure{#1} #3\par\egroup\endinsert}

\def\beneathrel#1\under#2{\mathrel{\mathop{#2}\limits_{#1}}}

\def\refto#1{$^{#1}$}		

\def\references			
  {\head{References}		
   \beginparmode
   \frenchspacing \parindent=0pt \leftskip=1truecm
   \parskip=8pt plus 3pt \everypar{\hangindent=\parindent}}

\gdef\refis#1{\item{#1.\ }}			

\gdef\journal#1, #2, #3, 1#4#5#6{		
    {\sl #1~}{\bf #2}, #3 (1#4#5#6)}		

\def\pr{\journal Phys. Rev., }

\def\prb{\journal Phys. Rev. B, }

\def\prl{\journal Phys. Rev. Lett., }

\def\np{\journal Nucl. Phys., }

\def\pl{\journal Phys. Lett., }

\def\endreferences{\body}

\def\figurecaptions		
  {\endpage
   \beginparmode
   \head{Figure Captions}
}

\def\endpage			
  {\vfill\eject}

\def\endpaper			
  {\endmode\vfill\supereject}


\def\heading				
  {\vskip 0.5truein plus 0.1truein	
   \beginparmode \def\\{\par} \parskip=0pt \singlespace \raggedcenter}

\def\subheading				
  {\vskip 0.25truein plus 0.1truein	
   \beginlinemode \singlespace \parskip=0pt \def\\{\par}\raggedcenter}

\def\tag#1$${\eqno(#1)$$}

\def\align#1$${\eqalign{#1}$$}

\def\aligntag#1$${\gdef\tag##1\\{&(##1)\cr}\eqalignno{#1\\}$$
  \gdef\tag##1$${\eqno(##1)$$}}

\def\endaligntag{}

\def\overset #1\to#2{{\mathop{#2}\limits^{#1}}}
\def\underset#1\to#2{{\let\next=#1\mathpalette\undersetpalette#2}}
\def\undersetpalette#1#2{\vtop{\baselineskip0pt
\ialign{$\mathsurround=0pt #1\hfil##\hfil$\crcr#2\crcr\next\crcr}}}


\def\ref#1{Ref.~#1}			
\def\Ref#1{Ref.~#1}			
\def\[#1]{[\cite{#1}]}
\def\cite#1{{#1}}
\def\(#1){(\call{#1})}
\def\call#1{{#1}}
\def\taghead#1{}
\def\frac#1#2{{#1 \over #2}}

\def\12{{1\over2}}

\def\etal{{\it et al.\ }}

\def\sla{\raise.15ex\hbox{$/$}\kern-.57em}
\def\leaderfill{\leaders\hbox to 1em{\hss.\hss}\hfill}
\def\twiddle{\lower.9ex\rlap{$\kern-.1em\scriptstyle\sim$}}
\def\bigtwiddle{\lower1.ex\rlap{$\sim$}}
\def\gtwid{\mathrel{\raise.3ex\hbox{$>$\kern-.75em\lower1ex\hbox{$\sim$}}}}
\def\ltwid{\mathrel{\raise.3ex\hbox{$<$\kern-.75em\lower1ex\hbox{$\sim$}}}}
\def\square{\kern1pt\vbox{\hrule height 1.2pt\hbox{\vrule width 1.2pt\hskip 3pt
   \vbox{\vskip 6pt}\hskip 3pt\vrule width 0.6pt}\hrule height 0.6pt}\kern1pt}
\def\tdot#1{\mathord{\mathop{#1}\limits^{\kern2pt\ldots}}}
\def\happyface{%
$\bigcirc\rlap{\lower0.3ex\hbox{$\kern-0.83em\scriptscriptstyle\smile$}%
\raise0.4ex\hbox{$\kern-0.6em\scriptstyle\cdot\cdot$}}$}
\def\sadface{%
$\bigcirc\rlap{\lower0.25ex\hbox{$\kern-0.83em\scriptscriptstyle\frown$}%
\raise0.43ex\hbox{$\kern-0.6em\scriptstyle\cdot\cdot$}}$}

\def\pmb#1{\setbox0=\hbox{#1}%
  \kern-.025em\copy0\kern-\wd0
  \kern  .05em\copy0\kern-\wd0
  \kern-.025em\raise.0433em\box0 }


\catcode`@=11
\newwrite\tocfile\openout\tocfile=\jobname_toc
\newlinechar=`^^J
\write\tocfile{\string\input\space jnl^^J
  \string\pageno=-1\string\firstpageno=-1000\string\singlespace
  \string\null\string\vfill\string\centerline{TABLE OF CONTENTS}^^J
  \string\vskip 0.5 truein\string\rightline{\string\underbar{Page}}\smallskip}

\def\tocitem#1{
  \t@cskip{#1}\bigskip}
\def\tocitemitem#1{
  \t@cskip{\quad#1}\medskip}
\def\tocitemitemitem#1{
  \t@cskip{\qquad#1}\smallskip}
\def\tocitemall#1{
  \xdef#1##1{#1{##1}\noexpand\tocitem{##1}}}
\def\tocitemitemall#1{
  \xdef#1##1{#1{##1}\noexpand\tocitemitem{##1}}}
\def\tocitemitemitemall#1{
  \xdef#1##1{#1{##1}\noexpand\tocitemitemitem{##1}}}

\def\t@cskip#1#2{
  \write\tocfile{\string#2\string\line{^^J
  #1\string\leaderfill\space\number\folio}}}

%

\def\t@cproduce{
  \write\tocfile{\string\vfill\string\vfill\string\supereject\string\end}
  \closeout\tocfile
  \immediate\write16{Table of Contents written on []\jobname_TOC.TEX}}


\let\t@cend=\endpaper\def\endpaper{\t@cproduce\t@cend}

\catcode`@=12

\tocitemall\head		

\catcode`@=11
\newcount\r@fcount \r@fcount=0
\def\refreset{\global\r@fcount=0}
\newcount\r@fcurr
\immediate\newwrite\reffile
\newif\ifr@ffile\r@ffilefalse
\def\w@rnwrite#1{\ifr@ffile\immediate\write\reffile{#1}\fi\message{#1}}

\def\writer@f#1>>{}
\def\referencefile{
  \r@ffiletrue\immediate\openout\reffile=\jobname.ref%
  \def\writer@f##1>>{\ifr@ffile\immediate\write\reffile%
    {\noexpand\refis{##1} = \csname r@fnum##1\endcsname = %
     \expandafter\expandafter\expandafter\strip@t\expandafter%
     \meaning\csname r@ftext\csname r@fnum##1\endcsname\endcsname}\fi}%
  \def\strip@t##1>>{}}

\def\citeall#1{\xdef#1##1{#1{\noexpand\cite{##1}}}}
\def\cite#1{\each@rg\citer@nge{#1}}	

\def\each@rg#1#2{{\let\thecsname=#1\expandafter\first@rg#2,\end,}}
\def\first@rg#1,{\thecsname{#1}\apply@rg}	
\def\apply@rg#1,{\ifx\end#1\let\next=\relax
\else,\thecsname{#1}\let\next=\apply@rg\fi\next}

\def\citer@nge#1{\citedor@nge#1-\end-}	
\def\citer@ngeat#1\end-{#1}
\def\citedor@nge#1-#2-{\ifx\end#2\r@featspace#1 
  \else\citel@@p{#1}{#2}\citer@ngeat\fi}	
\def\citel@@p#1#2{\ifnum#1>#2{\errmessage{Reference range #1-#2\space is bad.}%
    \errhelp{If you cite a series of references by the notation M-N, then M and
    N must be integers, and N must be greater than or equal to M.}}\else%
 {\count0=#1\count1=#2\advance\count1
by1\relax\expandafter\r@fcite\the\count0,%
  \loop\advance\count0 by1\relax
    \ifnum\count0<\count1,\expandafter\r@fcite\the\count0,%
  \repeat}\fi}

\def\r@featspace#1#2 {\r@fcite#1#2,}	
\def\r@fcite#1,{\ifuncit@d{#1}
    \newr@f{#1}%
    \expandafter\gdef\csname r@ftext\number\r@fcount\endcsname%
                     {\message{Reference #1 to be supplied.}%
                      \writer@f#1>>#1 to be supplied.\par}%
 \fi%
 \csname r@fnum#1\endcsname}
\def\ifuncit@d#1{\expandafter\ifx\csname r@fnum#1\endcsname\relax}%
\def\newr@f#1{\global\advance\r@fcount by1%
    \expandafter\xdef\csname r@fnum#1\endcsname{\number\r@fcount}}

\let\r@fis=\refis			
\def\refis#1#2#3\par{\ifuncit@d{#1}
   \newr@f{#1}%
   \w@rnwrite{Reference #1=\number\r@fcount\space is not cited up to now.}\fi%
  \expandafter\gdef\csname r@ftext\csname r@fnum#1\endcsname\endcsname%
  {\writer@f#1>>#2#3\par}}

\def\ignoreuncited{
   \def\refis##1##2##3\par{\ifuncit@d{##1}%
     \else\expandafter\gdef\csname r@ftext\csname
r@fnum##1\endcsname\endcsname%
     {\writer@f##1>>##2##3\par}\fi}}

\def\r@ferr{\endreferences\errmessage{I was expecting to see
\noexpand\endreferences before now;  I have inserted it here.}}
\let\r@ferences=\references
\def\references{\r@ferences\def\endmode{\r@ferr\par\endgroup}}

\let\endr@ferences=\endreferences
\def\endreferences{\r@fcurr=0
  {\loop\ifnum\r@fcurr<\r@fcount
    \advance\r@fcurr by 1\relax\expandafter\r@fis\expandafter{\number\r@fcurr}%
    \csname r@ftext\number\r@fcurr\endcsname%
  \repeat}\gdef\r@ferr{}\global\r@fcount=0\endr@ferences}


\let\r@fend=\endpaper\gdef\endpaper{\ifr@ffile
\immediate\write16{Cross References written on []\jobname.REF.}\fi\r@fend}

\catcode`@=12

\citeall\refto		
\citeall\ref		%
\citeall\Ref		%

\catcode`@=11
\newcount\tagnumber\tagnumber=0

\immediate\newwrite\eqnfile
\newif\if@qnfile\@qnfilefalse
\def\write@qn#1{}
\def\writenew@qn#1{}
\def\w@rnwrite#1{\write@qn{#1}\message{#1}}
\def\@rrwrite#1{\write@qn{#1}\errmessage{#1}}

\def\taghead#1{\gdef\t@ghead{#1}\global\tagnumber=0}
\def\t@ghead{}

\expandafter\def\csname @qnnum-3\endcsname
  {{\t@ghead\advance\tagnumber by -3\relax\number\tagnumber}}
\expandafter\def\csname @qnnum-2\endcsname
  {{\t@ghead\advance\tagnumber by -2\relax\number\tagnumber}}
\expandafter\def\csname @qnnum-1\endcsname
  {{\t@ghead\advance\tagnumber by -1\relax\number\tagnumber}}
\expandafter\def\csname @qnnum0\endcsname
  {\t@ghead\number\tagnumber}
\expandafter\def\csname @qnnum+1\endcsname
  {{\t@ghead\advance\tagnumber by 1\relax\number\tagnumber}}
\expandafter\def\csname @qnnum+2\endcsname
  {{\t@ghead\advance\tagnumber by 2\relax\number\tagnumber}}
\expandafter\def\csname @qnnum+3\endcsname
  {{\t@ghead\advance\tagnumber by 3\relax\number\tagnumber}}

\def\equationfile{%
  \@qnfiletrue\immediate\openout\eqnfile=\jobname.eqn%
  \def\write@qn##1{\if@qnfile\immediate\write\eqnfile{##1}\fi}
  \def\writenew@qn##1{\if@qnfile\immediate\write\eqnfile
    {\noexpand\tag{##1} = (\t@ghead\number\tagnumber)}\fi}
}

\def\callall#1{\xdef#1##1{#1{\noexpand\call{##1}}}}
\def\call#1{$\rm\each@rg\callr@nge{#1}$}

\def\each@rg#1#2{{\let\thecsname=#1\expandafter\first@rg#2,\end,}}
\def\first@rg#1,{\thecsname{#1}\apply@rg}
\def\apply@rg#1,{\ifx\end#1\let\next=\relax%
\else,\thecsname{#1}\let\next=\apply@rg\fi\next}

\def\callr@nge#1{\calldor@nge#1-\end-}
\def\callr@ngeat#1\end-{#1}
\def\calldor@nge#1-#2-{\ifx\end#2\@qneatspace#1 %
  \else\calll@@p{#1}{#2}\callr@ngeat\fi}
\def\calll@@p#1#2{\ifnum#1>#2{\@rrwrite{Equation range #1-#2\space is bad.}
\errhelp{If you call a series of equations by the notation M-N, then M and
N must be integers, and N must be greater than or equal to M.}}\else%
 {\count0=#1\count1=#2\advance\count1
by1\relax\expandafter\@qncall\the\count0,%
  \loop\advance\count0 by1\relax%
    \ifnum\count0<\count1,\expandafter\@qncall\the\count0,%
  \repeat}\fi}

\def\@qneatspace#1#2 {\@qncall#1#2,}
\def\@qncall#1,{\ifunc@lled{#1}{\def\next{#1}\ifx\next\empty\else
  \w@rnwrite{Equation number \noexpand\(>>#1<<) has not been defined yet.}
  >>#1<<\fi}\else\csname @qnnum#1\endcsname\fi}

\let\eqnono=\eqno
\def\eqno(#1){\tag#1}
\def\tag#1$${\eqnono(\displayt@g#1 )$$}

\def\aligntag#1\endaligntag
  $${\gdef\tag##1\\{&(##1 )\cr}\eqalignno{#1\\}$$
  \gdef\tag##1$${\eqnono(\displayt@g##1 )$$}}

\def\eqalignno#1{\displ@y \tabskip\centering
  \halign to\displaywidth{\hfil$\displaystyle{##}$\tabskip\z@skip
    &$\displaystyle{{}##}$\hfil\tabskip\centering
    &\llap{$\displayt@gpar##$}\tabskip\z@skip\crcr
    #1\crcr}}

\def\displayt@gpar(#1){(\displayt@g#1 )}

\def\displayt@g#1 {\rm\ifunc@lled{#1}\global\advance\tagnumber by1
        {\def\next{#1}\ifx\next\empty\else\expandafter
        \xdef\csname @qnnum#1\endcsname{\t@ghead\number\tagnumber}\fi}%
  \writenew@qn{#1}\t@ghead\number\tagnumber\else
        {\edef\next{\t@ghead\number\tagnumber}%
        \expandafter\ifx\csname @qnnum#1\endcsname\next\else
        \w@rnwrite{Equation \noexpand\tag{#1} is a duplicate number.}\fi}%
  \csname @qnnum#1\endcsname\fi}

\def\ifunc@lled#1{\expandafter\ifx\csname @qnnum#1\endcsname\relax}

\let\@qnend=\end\gdef\end{\if@qnfile
\immediate\write16{Equation numbers written on []\jobname.EQN.}\fi\@qnend}

\catcode`@=12


%
\newbox\hdbox%
\newcount\hdrows%
\newcount\multispancount%
\newcount\ncase%
\newcount\ncols
\newcount\nrows%
\newcount\nspan%
\newcount\ntemp%
\newdimen\hdsize%
\newdimen\newhdsize%
\newdimen\parasize%
\newdimen\spreadwidth%
\newdimen\thicksize%
\newdimen\thinsize%
\newdimen\tablewidth%
\newif\ifcentertables%
\newif\ifendsize%
\newif\iffirstrow%
\newif\iftableinfo%
\newtoks\dbt%
\newtoks\hdtks%
\newtoks\savetks%
\newtoks\tableLETtokens%
\newtoks\tabletokens%
\newtoks\widthspec%
%
%
%
%
\tableinfotrue%
\catcode`\@=11
%
%
\def\tstrut{\vrule height3.1ex depth1.2ex width0pt}%
\def\and{\char`\&}
\def\tablerule{\noalign{\hrule height\thinsize depth0pt}}%
\thicksize=1.5pt
\thinsize=0.6pt
\def\thickrule{\noalign{\hrule height\thicksize depth0pt}}%
\def\ctr#1{\hfil\quad\ #1\quad\hfil}%
%
%
%
%
\tablewidth=-\maxdimen%
\spreadwidth=-\maxdimen%
\def\tabskipglue{0pt plus 1fil minus 1fil}%
%
%
\centertablestrue%
%
%
%
%
\parasize=4in%
\gdef\ARGS{########}
\gdef\headerARGS{####}
\def\@mpersand{&}
{\catcode`\|=13
\gdef\letbarzero{\let|0}
\gdef\letbartab{\def|{&&}}%
\gdef\letvbbar{\let\vb|}%
}
{\catcode`\&=4
\def\ampskip{&\omit\hfil&}
\catcode`\&=13
\let&0
\xdef\letampskip{\def&{\ampskip}}%
\gdef\letnovbamp{\let\novb&\let\tab&}
}
\def\begintable{
   \begingroup%
   \catcode`\|=13\letbartab\letvbbar%
   \catcode`\&=13\letampskip\letnovbamp%
   \def\multispan##1{
      \omit \mscount##1%
      \multiply\mscount\tw@\advance\mscount\m@ne%
      \loop\ifnum\mscount>\@ne \sp@n\repeat%
   }
   \def\|{%
      &\omit\widevline&%
   }%
   \ruledtable
}
\long\def\ruledtable#1\endtable{%
%
%
%
   \offinterlineskip
   \tabskip 0pt
   \def\widevline{\vrule width\thicksize}
   \def\endrow{\@mpersand\omit\hfil\crnorm\@mpersand}%
   \def\crthick{\@mpersand\crnorm\thickrule\@mpersand}%
   \def\crthickneg##1{\@mpersand\crnorm\thickrule
          \noalign{{\skip0=##1\vskip-\skip0}}\@mpersand}%
   \def\crnorule{\@mpersand\crnorm\@mpersand}%
   \def\crnoruleneg##1{\@mpersand\crnorm
          \noalign{{\skip0=##1\vskip-\skip0}}\@mpersand}%
   \let\nr=\crnorule
   \def\endtable{\@mpersand\crnorm\thickrule}%
   \let\crnorm=\cr
%
%
   \edef\cr{\@mpersand\crnorm\tablerule\@mpersand}%
   \def\crneg##1{\@mpersand\crnorm\tablerule
          \noalign{{\skip0=##1\vskip-\skip0}}\@mpersand}%
   \let\ctneg=\crthickneg
   \let\nrneg=\crnoruleneg
   \the\tableLETtokens
%
%
   \tabletokens={&#1}
%
%
   \countROWS\tabletokens\into\nrows%
   \countCOLS\tabletokens\into\ncols%
%
%
   \advance\ncols by -1%
   \divide\ncols by 2%
   \advance\nrows by 1%
%
%
   \iftableinfo %
      \immediate\write16{[Nrows=\the\nrows, Ncols=\the\ncols]}%
   \fi%
%
%
   \ifcentertables
      \ifhmode \par\fi
      \line{
      \hss
   \else %
      \hbox{%
   \fi
      \vbox{%
         \makePREAMBLE{\the\ncols}
         \edef\next{\preamble}
         \let\preamble=\next
         \makeTABLE{\preamble}{\tabletokens}
      }
      \ifcentertables \hss}\else }\fi
   \endgroup
   \tablewidth=-\maxdimen
   \spreadwidth=-\maxdimen
}
\def\makeTABLE#1#2{
   {
   \let\ifmath0
   \let\header0
   \let\multispan0
%
%
   \ncase=0%
   \ifdim\tablewidth>-\maxdimen \ncase=1\fi%
   \ifdim\spreadwidth>-\maxdimen \ncase=2\fi%
   \relax
%
   \ifcase\ncase %
      \widthspec={}%
   \or %
      \widthspec=\expandafter{\expandafter t\expandafter o%
                 \the\tablewidth}%
   \else %
      \widthspec=\expandafter{\expandafter s\expandafter p\expandafter r%
                 \expandafter e\expandafter a\expandafter d%
                 \the\spreadwidth}%
   \fi %
   \xdef\next{
      \halign\the\widthspec{%
      #1
      \noalign{\hrule height\thicksize depth0pt}
      \the#2\endtable
%
      }
   }
   }
   \next
}
\def\makePREAMBLE#1{
   \ncols=#1
   \begingroup
   \let\ARGS=0
   \edef\xtp{\widevline\ARGS\tabskip\tabskipglue%
   &\ctr{\ARGS}\tstrut}
   \advance\ncols by -1
   \loop
      \ifnum\ncols>0 %
      \advance\ncols by -1%
      \edef\xtp{\xtp&\vrule width\thinsize\ARGS&\ctr{\ARGS}}%
   \repeat
   \xdef\preamble{\xtp&\widevline\ARGS\tabskip0pt%
   \crnorm}
   \endgroup
}
\def\countROWS#1\into#2{
   \let\countREGISTER=#2%
   \countREGISTER=0%
   \expandafter\ROWcount\the#1\endcount%
}%
\def\ROWcount{%
   \afterassignment\subROWcount\let\next= %
}%
\def\subROWcount{%
   \ifx\next\endcount %
      \let\next=\relax%
   \else%
      \ncase=0%
      \ifx\next\cr %
         \global\advance\countREGISTER by 1%
         \ncase=0%
      \fi%
      \ifx\next\endrow %
         \global\advance\countREGISTER by 1%
         \ncase=0%
      \fi%
      \ifx\next\crthick %
         \global\advance\countREGISTER by 1%
         \ncase=0%
      \fi%
      \ifx\next\crnorule %
         \global\advance\countREGISTER by 1%
         \ncase=0%
      \fi%
      \ifx\next\crthickneg %
         \global\advance\countREGISTER by 1%
         \ncase=0%
      \fi%
      \ifx\next\crnoruleneg %
         \global\advance\countREGISTER by 1%
         \ncase=0%
      \fi%
      \ifx\next\crneg %
         \global\advance\countREGISTER by 1%
         \ncase=0%
      \fi%
      \ifx\next\header %
         \ncase=1%
      \fi%
      \relax%
      \ifcase\ncase %
         \let\next\ROWcount%
      \or %
         \let\next\argROWskip%
      \else %
      \fi%
   \fi%
   \next%
}
\def\counthdROWS#1\into#2{%
\dvr{10}%
   \let\countREGISTER=#2%
   \countREGISTER=0%
\dvr{11}%
\dvr{13}%
   \expandafter\hdROWcount\the#1\endcount%
\dvr{12}%
}%
\def\hdROWcount{%
   \afterassignment\subhdROWcount\let\next= %
}%
\def\subhdROWcount{%
   \ifx\next\endcount %
      \let\next=\relax%
   \else%
      \ncase=0%
      \ifx\next\cr %
         \global\advance\countREGISTER by 1%
         \ncase=0%
      \fi%
      \ifx\next\endrow %
         \global\advance\countREGISTER by 1%
         \ncase=0%
      \fi%
      \ifx\next\crthick %
         \global\advance\countREGISTER by 1%
         \ncase=0%
      \fi%
      \ifx\next\crnorule %
         \global\advance\countREGISTER by 1%
         \ncase=0%
      \fi%
      \ifx\next\header %
         \ncase=1%
      \fi%
\relax%
      \ifcase\ncase %
         \let\next\hdROWcount%
      \or%
         \let\next\arghdROWskip%
      \else %
      \fi%
   \fi%
   \next%
}%
{\catcode`\|=13\letbartab
\gdef\countCOLS#1\into#2{%
   \let\countREGISTER=#2%
   \global\countREGISTER=0%
   \global\multispancount=0%
   \global\firstrowtrue
   \expandafter\COLcount\the#1\endcount%
   \global\advance\countREGISTER by 3%
   \global\advance\countREGISTER by -\multispancount
}%
\gdef\COLcount{%
   \afterassignment\subCOLcount\let\next= %
}%
{\catcode`\&=13%
\gdef\subCOLcount{%
   \ifx\next\endcount %
      \let\next=\relax%
   \else%
      \ncase=0%
      \iffirstrow
         \ifx\next& %
            \global\advance\countREGISTER by 2%
            \ncase=0%
         \fi%
         \ifx\next\span %
            \global\advance\countREGISTER by 1%
            \ncase=0%
         \fi%
         \ifx\next| %
            \global\advance\countREGISTER by 2%
            \ncase=0%
         \fi
         \ifx\next\|
            \global\advance\countREGISTER by 2%
            \ncase=0%
         \fi
         \ifx\next\multispan
            \ncase=1%
            \global\advance\multispancount by 1%
         \fi
         \ifx\next\header
            \ncase=2%
         \fi
         \ifx\next\cr       \global\firstrowfalse \fi
         \ifx\next\endrow   \global\firstrowfalse \fi
         \ifx\next\crthick  \global\firstrowfalse \fi
         \ifx\next\crnorule \global\firstrowfalse \fi
         \ifx\next\crnoruleneg \global\firstrowfalse \fi
         \ifx\next\crthickneg  \global\firstrowfalse \fi
         \ifx\next\crneg       \global\firstrowfalse \fi
      \fi
\relax
      \ifcase\ncase %
         \let\next\COLcount%
      \or %
         \let\next\spancount%
      \or %
         \let\next\argCOLskip%
      \else %
      \fi %
   \fi%
   \next%
}%
\gdef\argROWskip#1{%
   \let\next\ROWcount \next%
}
\gdef\arghdROWskip#1{%
   \let\next\ROWcount \next%
}
\gdef\argCOLskip#1{%
   \let\next\COLcount \next%
}
}
}
\def\spancount#1{
   \nspan=#1\multiply\nspan by 2\advance\nspan by -1%
   \global\advance \countREGISTER by \nspan
   \let\next\COLcount \next}%
\def\dvr#1{\relax}%
\def\header#1{%
\dvr{1}{\let\cr=\@mpersand%
\hdtks={#1}%
\counthdROWS\hdtks\into\hdrows%
\advance\hdrows by 1%
\ifnum\hdrows=0 \hdrows=1 \fi%
\dvr{5}\makehdPREAMBLE{\the\hdrows}%
\dvr{6}\getHDdimen{#1}%
{\parindent=0pt\hsize=\hdsize{\let\ifmath0%
\xdef\next{\valign{\headerpreamble #1\crnorm}}}\dvr{7}\next\dvr{8}%
}%
}\dvr{2}}
\def\makehdPREAMBLE#1{
\dvr{3}%
\hdrows=#1
{
\let\headerARGS=0%
\let\cr=\crnorm%
\edef\xtp{\vfil\hfil\hbox{\headerARGS}\hfil\vfil}%
\advance\hdrows by -1
\loop
\ifnum\hdrows>0%
\advance\hdrows by -1%
\edef\xtp{\xtp&\vfil\hfil\hbox{\headerARGS}\hfil\vfil}%
\repeat%
\xdef\headerpreamble{\xtp\crcr}%
}
\dvr{4}}
\def\getHDdimen#1{%
\hdsize=0pt%
\getsize#1\cr\end\cr%
}
\def\getsize#1\cr{%
\endsizefalse\savetks={#1}%
\expandafter\lookend\the\savetks\cr%
\relax \ifendsize \let\next\relax \else%
\setbox\hdbox=\hbox{#1}\newhdsize=1.0\wd\hdbox%
\ifdim\newhdsize>\hdsize \hdsize=\newhdsize \fi%
\let\next\getsize \fi%
\next%
}%
\def\lookend{\afterassignment\sublookend\let\looknext= }%
\def\sublookend{\relax%
\ifx\looknext\cr %
\let\looknext\relax \else %
   \relax
   \ifx\looknext\end \global\endsizetrue \fi%
   \let\looknext=\lookend%
    \fi \looknext%
}%
%
%
\def\tablelet#1{%
   \tableLETtokens=\expandafter{\the\tableLETtokens #1}%
}%
\catcode`\@=12
%


\def\CC{\hbox{C\kern -.58em {\raise .54ex \hbox{$\scriptscriptstyle |$}}
\kern-.55em {\raise .53ex \hbox{$\scriptscriptstyle |$}} }}
\def\RR{\hbox{I\kern-.2em\hbox{R}}}
\def\sRR{{\sl \hbox{I\kern-.2em\hbox{R}}}}
\def\sqr#1#2{{\vcenter{\vbox{\hrule height.#2pt\hbox{\vrule width.#2pt
height#1pt \kern#1pt\vrule width.#2pt}\hrule height.#2pt}}}}
\def\square{\mathchoice\sqr54\sqr54\sqr33\sqr23}






%
\def\LaTeX{{\rm L\kern-.36em\raise.3ex\hbox{\sc a}\kern-.15em
    T\kern-.1667em\lower.7ex\hbox{E}\kern-.125emX}}


\def\etal{{\it et al.}}



\def\slD{\raise.15ex\hbox{$/$}\kern-.53em\hbox{$D$}}
\def\slA{\raise.15ex\hbox{$/$}\kern-.53em\hbox{$A$}}
\def\dsl{\raise.15ex\hbox{$/$}\kern-.57em\hbox{$\Delta$}}
\def\slp{{\raise.15ex\hbox{$/$}\kern-.57em\hbox{$\partial$}}}
\def\nsl{\raise.15ex\hbox{$/$}\kern-.57em\hbox{$\nabla$}}
\def\sla{\raise.15ex\hbox{$/$}\kern-.57em\hbox{$\rightarrow$}}
\def\slla{\raise.15ex\hbox{$/$}\kern-.57em\hbox{$\lambda$}}
\def\slb{\raise.15ex\hbox{$/$}\kern-.57em\hbox{$b$}}
\def\slr{\raise.15ex\hbox{$/$}\kern-.57em\hbox{$r$}}
\def\lnp{\raise.15ex\hbox{$/$}\kern-.57em\hbox{$p$}}
\def\lnk{\raise.15ex\hbox{$/$}\kern-.57em\hbox{$k$}}
\def\lnK{\raise.15ex\hbox{$/$}\kern-.57em\hbox{$K$}}
\def\lnq{\raise.15ex\hbox{$/$}\kern-.57em\hbox{$q$}}
\def\nna{\raise.15ex\hbox{$/$}\kern-.57em\hbox{$a$}}



\def\de{{\delta}}
\def\eps{{\epsilon}}

\def\la{\lambda}

\def\La{{\Lambda}}

\def\munu{{\mu\nu}}



\def\cL{{\cal L}}





\def\gtwid{\raise.3ex\hbox{$>$\kern-.75em\lower1ex\hbox{$\sim$}}}
\def\ltwid{\raise.3ex\hbox{$<$\kern-.75em\lower1ex\hbox{$\sim$}}}
\def\12{{1\over2}}
\def\part{\partial}

\def\low#1{\lower.5ex\hbox{${}_#1$}}

\def\partt{\raise.15ex\hbox{$\widetilde$}{\kern-.37em\hbox{$\partial$}}}

\def\fulltriangle{{{{{{{{{\pmb{\triangle}\kern-.65em\bullet}\kern-.4em
{\raise1.2ex\hbox{.}}}
\kern-.4em{\raise1.0ex\hbox{.}}}\kern-.2em{\raise1.0ex\hbox{.}}}
\kern-.4em{\raise.1ex\hbox{.}}}\kern-.4em{\raise.2ex\hbox{.}}}
\kern-.2em{\raise.35ex\hbox{.}}}\kern.1em{\raise.2ex\hbox{.}} }}
                                                                \
\def\hexagon{{\tenpoint
\langle\kern-.1em{\raise.2cm\hbox{$\overline{\hskip.7em\relax}$}}
\kern-.7em{\lower.3ex\hbox{$\underline{\hskip.7em\relax}$}}\kern-.075em
 \rangle}}

\def\pentagon{{\tenpoint
\raise.5ex\hbox{$\widehat{\qquad}$}\kern-1.8em{\backslash
\kern-.1em{\lower.3ex\hbox{$\underline{\kern.75em}$}}\kern-.05em/} }}


\def\topppageno1{\global\footline={\hfil}\global\headline
={\ifnum\pageno<\firstpageno{\hfil}\else{\hss\twelverm --\ \folio
\ --\hss}\fi}}

\def\toppageno2{\global\footline={\hfil}\global\headline
={\ifnum\pageno<\firstpageno{\hfil}\else{\rightline{\hfill\hfill
\twelverm \ \folio
\ \hss}}\fi}}

\def\boxit#1{\vbox{\hrule\hbox{\vrule\kern3pt
  \vbox{\kern3pt#1\kern3pt}\kern3pt\vrule}\hrule}}